\documentclass[usenatbib]{mn2e}
\usepackage{natbib}
\usepackage{epsf}
\usepackage{graphicx}

\title[Global OIR--X-ray correlations in X-ray binaries]{Global optical/infrared--X-ray correlations in X-ray binaries: quantifying disc and jet contributions}
\author[D. M. Russell et al.]
{D. M. Russell$^{1}$\thanks{Email: davidr@phys.soton.ac.uk},
R. P. Fender$^{1}$, R. I. Hynes$^{2}$, C. Brocksopp$^{3}$, J. Homan$^{4}$,
\newauthor
P. G. Jonker$^{5,6,7}$, M. M. Buxton$^{8}$
\\
$^1$School of Physics \& Astronomy, University of Southampton, Highfield,
Southampton, SO17 1BJ, UK\\
$^2$Department of Physics and Astronomy, 202 Nicholson Hall, Louisiana State University, Baton Rouge, LA 70803, USA\\
$^3$Mullard Space Science Laboratory, Holmbury St Mary, Dorking, Surrey, RH5 6NT, UK\\
$^4$Massachusetts Institute of Technology, Kavli Institute for Astrophysics and Space Research, 70 Vassar Street, Cambridge, MA 02139, USA\\
$^5$SRON National Institute for Space Research, Corbonnelaan 2, 3584 CA Utrecht, Netherlands\\
$^6$Harvard-Smithsonian Center for Astrophysics, 60 Garden Street, MS 83, Cambridge, MA 02138, USA\\
$^7$Astronomical Institute, Utrecht University, P.O.Box 80000, 3508 TA, Utrecht, The Netherlands\\
$^8$Astronomy Department, Yale University, P.O. Box 208101, New Haven, CT 06520-8101, USA\\
}

\begin{document}
\maketitle

\begin{abstract}

The optical/near-infrared (OIR) region of the spectra of low-mass
X-ray binaries appears to lie at the intersection of a variety of
different emission processes. In this paper we present
quasi-simultaneous OIR--X-ray observations of 33 XBs in an attempt to
estimate the contributions of various emission processes in these
sources, as a function of X-ray state and luminosity. A global
correlation is found between OIR and X-ray luminosity for low-mass
black hole candidate XBs (BHXBs) in the hard X-ray state, of the form
$L_{\rm OIR}\propto L_{\rm X}^{0.6}$. This correlation holds over 8
orders of magnitude in $L_{\rm X}$ and includes data from BHXBs in
quiescence and at large distances (LMC and M31). A similar
correlation is found in low-mass neutron star XBs (NSXBs) in the hard
state. For BHXBs in the soft state, all the near-infrared (NIR) and
some of the optical emission is suppressed below the correlation, a
behaviour indicative of the jet switching off/on in transition to/from
the soft state. We compare these relations to theoretical models of a
number of emission processes. We find that X-ray reprocessing in the
disc and emission from the jets both predict a slope close to 0.6 for
BHXBs, and both contribute to the OIR in BHXBs in the hard state, the
jets producing $\sim90$ percent of the NIR emission at high
luminosities. X-ray reprocessing dominates the OIR in NSXBs in the
hard state, with possible contributions from the jets (only at high
luminosity) and the viscously heated disc. We also show that the
optically thick jet spectrum of BHXBs extends to near the $K$-band.
OIR spectral energy distributions of 15 BHXBs help us to confirm these
interpretations. We present a prediction of the $L_{\rm
OIR}$--$L_{\rm X}$ behaviour of a BHXB outburst that enters the soft
state, where the peak $L_{\rm OIR}$ in the hard state rise is greater
than in the hard state decline (the well known hysteretical
behaviour). In addition, it is possible to estimate the X-ray, OIR
and radio luminosity and the mass accretion rate in the hard state
quasi-simultaneously, from observations of just one of these
wavebands, since they are all linked through correlations. Finally,
we have discovered that the nature of the compact object, the mass of
the companion and the distance/reddening can be constrained by
quasi-simultaneous OIR and X-ray luminosities.
\end{abstract}


\section{Introduction}

X-ray binaries are binary systems in which a compact object -- either a black hole or a neutron star -- accretes matter from a companion star. The origin of the emission from these sources is known in some wavebands and not so well established in others. Radio emission is produced by the synchrotron process in collimated outflows \citep*[see e.g.][]{hjelet95}, whereas X-ray emission could originate e.g. directly from the hot inner accretion disc, from a Comptonising corona, from an advective flow, or from the jets \citep*{pout98,naraet98,mccoet00,market01,brocet04}.

The optical/near-infrared (optical/NIR; OIR) is perhaps the waveband for which the dominating emission processes in XBs are least known. OIR emission has been extensively studied in outburst and quiescence (for reviews see \citealt{vanpet95} and \citealt{charco06}; see also \citealt*{chenet97}). The complex variety of spectral, timing and luminosity properties of the OIR emission indicates that many processes may be contributing, each depending on a number of factors. In high mass X-ray binaries (HMXBs), the OIR light is largely dominated by the massive companion star in the system \citep*{vandet72,trevet80} with occasional additional contributions, for example from the reprocessing of X-rays. For low-mass neutron star X-ray binaries (NSXBs), there is strong evidence for a central X-ray source illuminating a disc that reprocesses the light to OIR wavelengths \citep*[see e.g.][]{mcclet79}.

The optical emission of a low-mass black hole candidate X-ray binary (BHXB) is generally thought to arise in the outer accretion disc as the result of X-ray reprocessing \citep*{cunn76,vrtiet90}, much like the NSXBs \citep*[e.g.][]{vanpet95}. Indeed, timing and spectral analysis in many cases has led to this conclusion \citep*[e.g.][]{wagnet91,callet95,obriet02,hyneet02a,hyne05}. However, reprocessed X-rays are often misleadingly \emph{assumed} to dominate the OIR light in BHXBs. Some observations of BHXBs point towards alternative physical processes contributing (and sometimes dominating) the OIR emission \citep[e.g.][]{homaet05}.

Intrinsic thermal emission from the viscously heated outer accretion disc is expected to contribute significant light in the optical, through UV to X-ray wavelengths \citep*{shaksu73,franet02}. OIR behaviour has revealed this process to play a role in some BHXBs \citep*[e.g.][]{kuul98,soriet99,brocet01a,brocet01b,homaet05}. Thermal emission from the companion star is observed in some low-mass XBs in quiescence \citep*[e.g.][]{oke77,bail92,oroset96,greeet01,mikoet05}. In the last decade evidence has been mounting for the flat optically thick spectrum of the jets to extend from the radio to the OIR regime (\citealt*{hanet92,fend01,corbet02,market03,chatet03}; \citealt{brocet04,buxtet04,homaet05}; for a review see \citealt{fend06}). Behaviour that is not consistent with intrinsic disc or reprocessed emission has in the past been attributed to e.g. magnetic loop reconnection in the outer disc \citep*[e.g.][]{zuriet03}, emission from a magnetically dominated compact corona \citep*[e.g.][]{merlet00} or emission from an advective region \citep[e.g.][]{shahet03}.

In BHXBs (both transient and persistent), properties of the emission in all wavebands are often related to changes in the X-ray spectrum. The two main X-ray spectral states are the \emph{hard} (or \emph{low/hard}) state, which is characterised by a hard power-law spectrum and strong variability and the \emph{soft} (or \emph{high/soft}; \emph{thermal--dominant}) state, where a thermal spectrum dominates with a power-law contribution (see \citealt*{mcclet06} for a review of X-ray states; see also \citealt*{homaet01,fendet04,homabe05}). The low luminosity `quiescent' state is likely to be an extension to the hard state (\citealt*{nara96,esinet97,mcclet03,fendet03}; \citealt{fendet04,gallet06}) but currently this is not universally accepted. Here, we treat quiescence as an extension to the hard state but we also show how our results differ when the quiescent data are removed. We hereafter class `optical' and `NIR' emission as that seen in the $BVRI$ ($\sim4400-7900\AA$) and $JHK$ ($\sim1.25-2.22\mu m$) wavebands, respectively.

\subsection{Towards a Unified Model for the OIR Behaviour in BHXBs}

Power-law correlations between OIR and X-ray luminosities are naturally expected from a number of emission processes. \cite{vanpet94} showed that the optical luminosity of an X-ray reprocessing accretion disc varies as $L_{\rm OPT}\propto T^2 \propto L_{\rm X}^{0.5}a$, where $T$ is the temperature and $a$ is the orbital separation of the system, and that this correlation has been observed in a selection of low-mass XBs. $L_{\rm OIR}$--$L_{\rm X}$ correlations are also expected when the OIR originates in the viscously heated disc as both X-ray and OIR are linked through the mass accretion rate (see Section 3.2).

In addition, OIR--X-ray correlations can be predicted if the OIR emission originates in the jets. Models of steady, compact jets demonstrate that the total jet power is related to the radio luminosity as $L_{\rm radio}\propto L_{\rm jet}^{1.4}$ \citep{blanko79,falcbi96,market01}. It was shown that the jet power is linearly proportional to the mass accretion rate in NSXBs and BHXBs in the hard state \citep*{falcbi96,miglfe06,kordet06} and the X-ray luminosity scales as $L_{\rm X}\propto$ \.m and $L_{\rm X}\propto$ \.m$^2$ for radiatively efficient and inefficient objects, respectively \citep[e.g][]{shaksu73,narayi95,kordet06}. The accretion in hard state BHXBs is found to be \emph{radiatively inefficient} (the majority of the liberated gravitational potential is carried in the flow and not radiated locally), where jet-dominated states can exist, whereas in NSXBs, the accretion is \emph{radiatively efficient}, and jet-dominated states are unlikely to exist \citep[see also][]{fendet03}. We therefore have:

\smallskip BHXBs: $L_{\rm radio}\propto L_{\rm jet}^{1.4}\propto $ \.m$^{1.4}\propto L_{\rm X}^{0.7}$

\smallskip NSXBs: $L_{\rm radio}\propto L_{\rm jet}^{1.4}\propto $ \.m$^{1.4}\propto L_{\rm X}^{1.4}$
\newline\newline
The correlation for BHXBs has been observed \citep*{corbet03,gallet03} and very recently, \cite*{miglfe06} have applied this technique to NSXBs and found $L_{\rm radio}\propto L_{\rm x}^{\ge1.4}$; which is also consistent with the above NSXB model. If the optically thick jet spectrum is indeed flat from the radio regime to OIR, we can expect the following correlations:

\smallskip BHXBs: $L_{\rm OIR}\propto L_{\rm radio}\propto L_{\rm X}^{0.7}$

\smallskip NSXBs: $L_{\rm OIR}\propto L_{\rm radio}\propto L_{\rm X}^{1.4}$
\newline\newline
\cite{homaet05} discovered a correlation between the quasi-simultaneous NIR (which was shown to originate in the jets) and X-ray fluxes for GX 339--4 in the hard state, with a slope $F_{\rm NIR}\propto F_{\rm X}^{0.53\pm 0.02}$ (3--100 keV). To date, no other sources have been tested for jet OIR emission using OIR--X-ray correlations.

It is now becoming clear that this profitable but simple technique of analysing the dependence of OIR and X-ray luminosities over many orders of magnitude, may prove fruitful for the understanding of the emission mechanisms involved. The unification of jet--X-ray state activity is now underway; a steady jet exists in the hard state, which is accelerated as the X-ray spectrum softens, and is finally quenched as it passes the `jet line' into the soft state \citep{fendet04}. A unification (if one exists) of the origins of OIR light from BHXBs in different spectral and luminosity states is desired to understand the behaviour of these systems. Furthermore, a measure of the level of OIR emission from jets may be used to constrain jet power estimates.

\section{methodology \& results}

For this work, we have collected OIR and X-ray data from a large number of BHXBs, NSXBs and HMXBs in order to find relations that may help determine the processes responsible for the OIR light in these systems. We apply the technique of testing the dependency of OIR luminosity with X-ray luminosity for the three types of XB and between different X-ray states in BHXBs, and attempt to identify the dominant emission mechanisms for BHXBs at a given luminosity and X-ray state.

A literature search for quasi-simultaneous (no more than $\sim$ 1 day between observations; for sources with outbursts $\leq$1 month in length we only use data with separations of $\leq$0.1 times the outburst length) X-ray and OIR fluxes from BHXBs was conducted. Where possible, tabulated fluxes or magnitudes were noted. In some cases we obtained data directly from the authors. We also made use of the \emph{DEXTER} applet provided by \emph{NASA ADS} to extract data from light curves where the data themselves were unattainable. For each source, the best estimates of its distance, optical extinction $A_{\rm V}$ and HI absorption column $N_{\rm H}$ were sought. Table 1 lists the properties of each BHXB for which data were obtained. We used non-simultaneous OIR--X-ray luminosities only in quiescence for some sources, and for these we have included errors that encompass all observed values of the quiescent flux in one of the two wavebands.

\begin{figure*}
\centering
\includegraphics[height=22cm,width=17cm,angle=0]{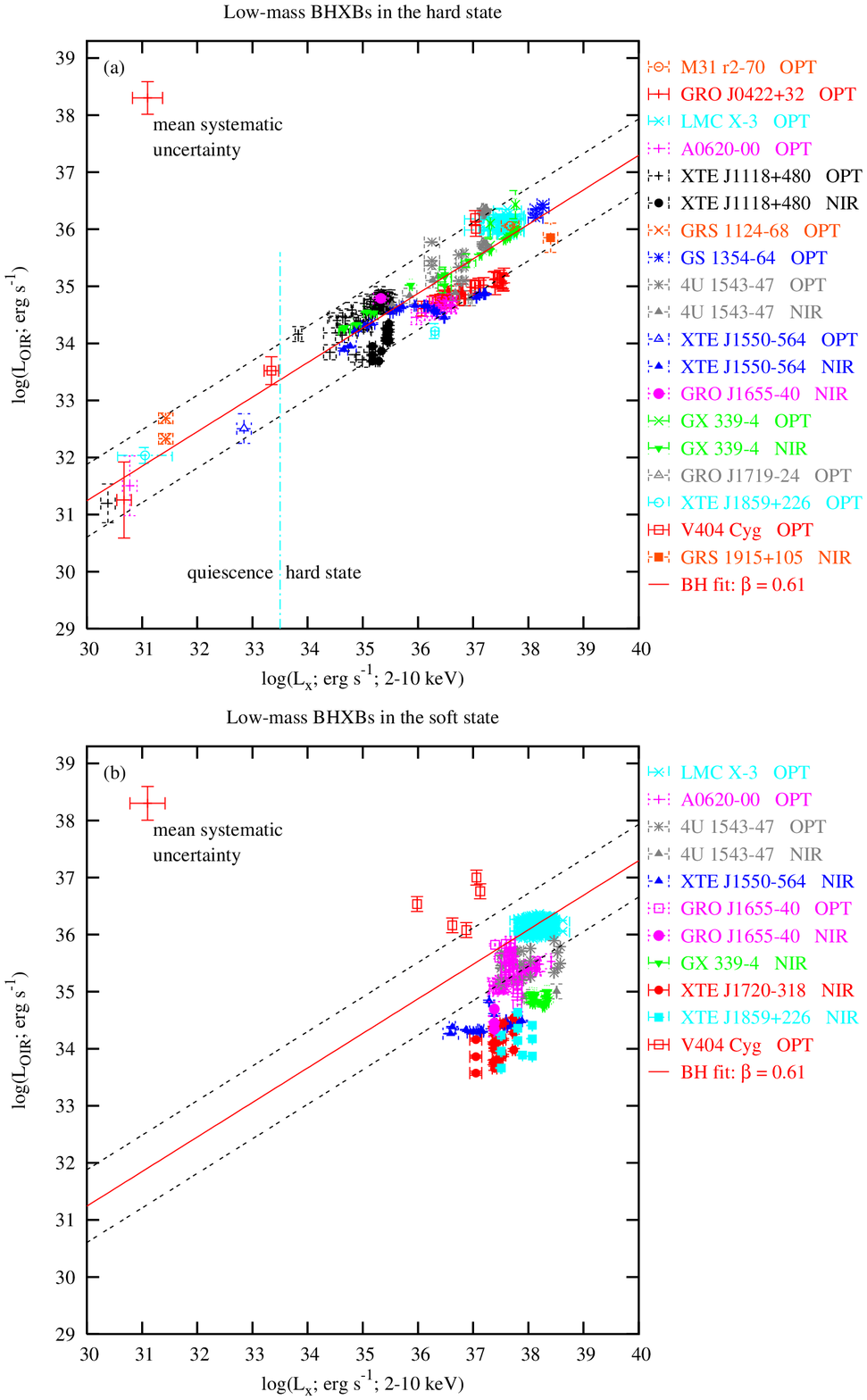}
\caption{OPT and NIR refer to the optical ($BVRI$) and NIR ($JHK$) wavebands, respectively. As a rule, filled symbols represent NIR data and unfilled symbols represent optical data. The mean 1$\sigma$ errors associated with each value (which include errors on distance and reddening) are shown top left. Panel (a): Quasi-simultaneous X-ray versus OIR luminosities for BHXBs in the hard X-ray state. The best power-law fit is $L_{\rm OIR} = 10^{13.1\pm 0.6} L_{\rm X}^{0.61\pm0.02}$. The black dashed lines represent the 1$\sigma$ uncertainties in the normalisation of the fit. Panel (b): As panel (a) but for BHXBs in the soft state (the BHXB hard state correlation is shown for comparison).
}
\label{fig:fig1}
\end{figure*}

\begin{figure*}
\includegraphics[height=22cm,width=17cm,angle=0]{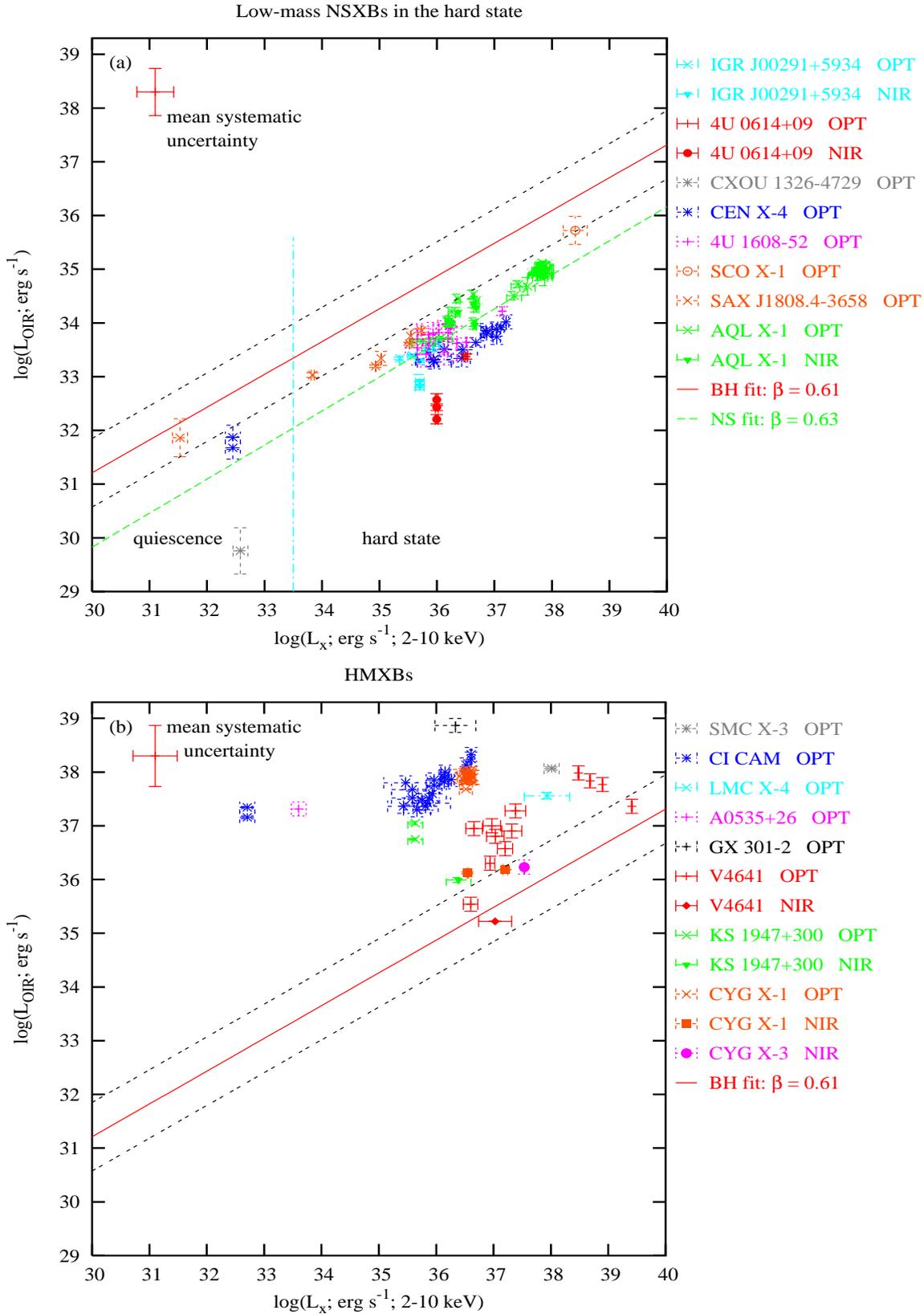}
\caption{OPT and NIR refer to the optical ($BVRI$) and NIR ($JHK$) wavebands, respectively. As a rule, filled symbols represent NIR data and unfilled symbols represent optical data. The mean 1$\sigma$ errors associated with each value (which include errors on distance and reddening) are shown top left. Panel (a): Quasi-simultaneous X-ray versus OIR luminosities for NSXBs in the hard X-ray state. The best power-law fit (green dashed line) is $L_{\rm OIR} = 10^{10.8\pm 1.4} L_{\rm X}^{0.63\pm0.04}$. The BHXB hard state correlation (red line with black dashed 1$\sigma$ uncertainties) is shown for comparison. Panel (b): As panel (a) but for HMXBs (black hole and neutron star XBs).
}
\label{fig:fig2}
\end{figure*}

\begin{table*}
\begin{center}
\caption{Properties and data collected for the 16 BHXBs.}
\begin{tabular}{lllllllll}
\hline
Source &Distance&Period &$M_{\rm co}$ &$M_{\rm cs}$ &$A_{\rm V}$, $N_{\rm H}$ / &$q_{\rm cs}$&$\Delta t$ /&Fluxes - \\
= alternative &/ kpc &/ hours&/ $M_\odot$&/ $M_\odot$&10$^{21} cm^{-2}$&(band, &days &data \\
name &(ref) &(ref) &(ref) &(ref) &(refs) &ref) & &references\\
(I)&(II)&(III)&(IV)&(V)&(VI)&(VII)&(VIII)&(IX)\\
\hline
M31 r2--70 &784$\pm$30 &192$^{+290}_{-120}$&-&-&1.0$\pm$0.3, &-&2.0&12\\
\vspace{2mm}
             &(1) &(12) & & &1.8$\pm$0.5 (12)& & & \\
GRO J0422+32 &2.49$\pm$0.30 &5.09&3.97$\pm$0.95&0.46$\pm$0.31&0.74$\pm$0.09,&0.76$^{+0.04}_{-0.46}$ &0.5&2,41,42\\
\vspace{2mm}
= V518 Per &(2) &(5) &(15) &(15) &1.6$\pm$0.4 (2,23) &(R, 2)& & \\
LMC X--3 &50$\pm$10 &$\sim$40.8&$\sim$9--10&$\sim$4--8&0.19$\pm$0.03,&-&1.0&43,44\\
\vspace{2mm}
             &(3,4) &(13) &(16) &(16) &0.32$^{+0.31}_{-0.07}$ (24--26) & & & \\
A0620--00 &1.2$\pm$0.4 &7.75&11.0$\pm$1.9&0.74$\pm$0.13&1.17$\pm$0.08, &0.58$^{+0.25}_{-0.22}$&1.0&33,45\\
\vspace{2mm}
= V616 Mon &(5) &(5) &(17) &(17) &2.4$^{+1.1}_{-1.0}$ (5,27)&(V, 33)& & \\
XTE J1118+480&1.71$\pm$0.05 &4.08&6.8$\pm$0.4&0.28$\pm$0.05&0.053$^{+0.027}_{-0.016}$, &0.55$^{+0.15}_{-0.28}$ &1.0&6,41--43,\\
\vspace{2mm}
= KV UMa &(6) &(5) &(15) &(15) &0.11$\pm$0.04 (6)&(R, 34)& &46--49 \\
GRS 1124--68 &5.5$\pm$1.0 &10.4&6.0$^{+1.5}_{-1.0}$&0.80$\pm$0.11&0.9$\pm$0.1 (5,28),&0.55$\pm$0.05&-&28,50--52\\
\vspace{1mm}
= GU Mus &(5) &(5) &(18) &(18) &1.58$^{+0.42}_{-0.58}$&\&(B--V, 35) & &\\
GS 1354--64 &$\geq$27 (7) &61.1&$>$7.83$\pm$0.50&$>$1.02$\pm$0.06&2.60$\pm$0.31, &-&1.0&43,53\\
\vspace{1mm}
= BW Cir &($\sim$33$\pm$6) &(7) &(15) &(15) &37.2$^{+14}_{-7}$ (7,29)& & & \\
4U 1543--47 &7.5$\pm$0.5 &26.8&9.4$\pm$1.0&2.45$\pm$0.15&1.55$\pm$0.15,&0.68$^{+ 0.11}_{-0.07}$&1.0&43,54,55\\
\vspace{1mm}
= IL Lup &(5) &(5) &(15) &(15) &4.3$\pm$0.2 (5) &(R, 36)& & \\
XTE J1550--564&5.3$\pm$2.3 &37.0&10.6$\pm$1.0&1.30$\pm$0.43&2.5$\pm$0.6,&0.7$\pm$0.1&1.0&19,56,57\\
\vspace{1mm}
= V381 Nor &(5) &(5) &(19) &(19) &8.7$\pm$2.1 (5)&(V, 19) & & \\
GRO J1655--40 &3.2$\pm$0.2 &62.9&7.02$\pm$0.22&2.35$\pm$0.14&3.7$\pm$0.3 (5), &$\sim$1.0&1.0&20,43,\\
\vspace{1mm}
= Nova Sco 1994&(5) &(5) &(20) &(20) &6.66$\pm$0.57 (20)&(B--K, 37)& &58--60 \\
GX 339--4 &8$^{+7.0}_{-1.0}$ &42.1&$\sim$5.8&$\sim$0.52&3.9$\pm$0.5,&$\leq$0.3&1.0&43,60--64\\
\vspace{1mm}
= V821 Ara &(8) &(5) &(21) &(21) &6$^{+0.9}_{-1.7}$ (5,30) &(B--K, 38) & & \\
GRO J1719--24 &2.4$\pm$0.4 &14.7&$\sim$4.9&$\sim$1.6&2.8$\pm$0.6,&-&0.5&41\\
\vspace{1mm}
= Nova Oph 1993&(9) &(14)&(9) &(9) &4$^{+0.0}_{-2.6}$ (9,31,32) & & & \\
XTE J1720--318&8$^{+7}_{-5}$ &-&-&-&6.9$\pm$0.1, &-&1.0&43,65\\
\vspace{1mm}
= INTEGRAL1 51&(10) & & & &12.4$\pm$0.2 (10$^{\ast}$)& & & \\
XTE J1859+226&6.3$\pm$1.7 &9.17&5--12&0.68--1.12&1.80$\pm$0.37,&0.59$\pm$0.04 &1.0&22,39,43,\\
\vspace{1mm}
= V406 Vul &(5) &(5) &(22)&(22) &8$\pm$2 (5) &(R, 39)& &66,67 \\
GRS 1915+105 &9.0$\pm$3.0 &816 &14.0$\pm$4.4&0.81$\pm$0.53&19.6$\pm$1.7,&-&1.0&68\\
\vspace{1mm}
= V1487 Aql &(11) &(5) &(15) &(15) &35$\pm$3 (11) & & & \\
V404 Cyg &4.0$^{+2.0}_{-1.2}$&155.28&10.0$\pm$2.0&0.65$\pm$0.25&3.65$\pm$0.35&0.87$\pm$0.03&0.5&69--71\\
\vspace{1mm}
= GS 2023+338&(5) &(5) &(15) &(15) &6.98$\pm$0.76 (5,27)&(R, 40) & & \\
\hline
\end{tabular}
\normalsize
\end{center}
Columns give:
(I) source names;
(II) distance estimate;
(III) orbital period of the system;
(IV) mass of the compact object (black hole here) in solar units;
(V) mass of the companion star in solar units;
(VI) interstellar reddening in $V$-band, and interstellar HI absorption column ($^{\ast}A_{\rm V}$ is estimated here from the relation $N_{\rm H} = 1.79 \times 10^{21} cm^{-2}A_{\rm V}$; \citealt{predet95});
(VII) the companion star OIR luminosity contribution in quiescence;
(VIII) The maximum time separation, $\Delta$t, between the OIR and X-ray observations defined as quasi-simultaneous;
(IX) References for the quasi-simultaneous OIR and X-ray fluxes collected.
References: see caption of Table 3.

\end{table*}

\begin{table*}
\begin{center}
\caption{Properties and data collected for the 8 NSXBs.}
\begin{tabular}{lllllllll}
\hline
Source &Distance&Period &$M_{\rm co}$ &$M_{\rm cs}$ &$A_{\rm V}$, $N_{\rm H}$ / &$q_{\rm cs}$&$\Delta t$ /&Fluxes - \\
= alternative &/ kpc &/ hours&/ $M_\odot$&/ $M_\odot$&10$^{21} cm^{-2}$&(band, &days &data \\
name &(ref) &(ref) &(ref) &(ref) &(refs) &ref) & &references\\
(I)&(II)&(III)&(IV)&(V)&(VI)&(VII)&(VIII)&(IX)\\
\hline
IGR J00291+5934 &4$^{+3}_{-0}$ &2.457&1.4&0.039--0.160&1.56$\pm$0.22 &-&1.0&97,98\\
\vspace{2mm}
                &(72) &(77) &(72)&(72) &2.8$\pm$0.4 (77$^{\ast}$) & & &\\
4U 0614+09 &3.0$^{+0.0}_{-2.5}$&0.25--0.33&1.4&$\leq$1.9 (80)&1.41$\pm$0.17 (78) &-&0.5&99,100\\
\vspace{2mm}
= V1055 Ori &(73) &(78) &(80)&($\sim$1.45) &2.99$\pm$0.01 (87)& & &\\
CXOU 132619.7 &5.0$\pm$0.5 &-&-&-&0.34$\pm$0.03 (88)&0.8$\pm$0.1 &-&74,88\\
\vspace{2mm}
--472910.8 &(74) & & & &0.9$\pm$0.1 (74) &(B, 88)& &\\
Cen X--4 &1.2$\pm$0.3 &15.1&1.3$\pm$0.8&0.31$\pm$0.27&0.31$\pm$0.16 (89)&0.75$\pm$0.05 (R), &0.5&79,95,\\
\vspace{2mm}
= V822 Cen &(75) &(79) &(15) &(85) &0.55$\pm$0.16 (90)&\multicolumn{2}{l}{0.70$\pm$0.05 (V, 95)}&101,102\\
4U 1608--52 &3.3$\pm$0.5 &$\sim$12.9&$\sim$1.4&$\sim$0.32&4.65$^{+3.25}_{-0.18}$ &-&1.0&43,81\\
\vspace{2mm}
= QX Nor &(5) &(5) &(81) &(81) &15$\pm$5 (14,81,91)& & &\\
Sco X--1 &2.8$\pm$0.3 &18.9&1.4 &0.42&0.70$\pm$0.23 &-&1.0&43,103\\
\vspace{2mm}
= V818 Sco &(5) &(5) &(82)&(82)&1.25$\pm$0.41 (14$^{\dagger}$)& & &\\
SAX J1808.4--3658&2.5$\pm$0.1 &2.0&$\geq$1.7 (83)&0.05--0.10&0.68$^{+0.37}_{-0.15}$ (86)&0.0$^{+0.3}_{-0.0}$&1.0&86,104,\\
\vspace{2mm}
= XTE J1808--369&(76) &(5)&($\sim$1.7) &(86) &0.11$\pm$0.03 (92) &(V, 96)& &105\\
Aql X--1 &5.15$\pm$0.75 &19.0&$\sim$1.4&$\sim$0.6&1.55$\pm$0.31 (93) &-&0.5&43,\\
\vspace{2mm}
= V1333 Aql &(5) &(5) &(84) &(84) &4.0$^{+3.8}_{-3.2}$ (94)& & &106--112\\
\hline
\end{tabular}
\normalsize
\end{center}
Columns give:
(I) source names;
(II) distance estimate;
(III) orbital period of the system;
(IV) mass of the compact object (neutron star here) in solar units;
(V) mass of the companion star in solar units;
(VI) interstellar reddening in $V$-band, and interstellar HI absorption column ($^{\ast}A_{\rm V}$ and $^{\dagger}N_{\rm H}$ are estimated here from the relation $N_{\rm H} = 1.79 \times 10^{21} cm^{-2}A_{\rm V}$; \citealt{predet95});
(VII) the companion star OIR luminosity contribution in quiescence;
(VIII) The maximum time separation, $\Delta t$, between the OIR and X-ray observations defined as quasi-simultaneous;
(IX) References for the quasi-simultaneous OIR and X-ray fluxes collected.
References: see caption of Table 3.

\end{table*}

\begin{table*}
\begin{center}
\caption{Properties and data collected for the 9 HMXBs.}
\begin{tabular}{lllllllll}
\hline
Source = alternative name&Compact&Distance &$A_{\rm V}$ (ref)&$N_{\rm H}$ / 10$^{21} cm^{-2}$&$\Delta t$&Fluxes - \\
                         &object &/ kpc (ref)& &(ref) &/ days &data references\\
(I)&(II)&(III)&(IV)&(V)&(VI)&(VII)\\
\hline
\rule[0mm]{0mm}{4mm}
SMC X--3 &NS &58.1$\pm$5.6 (113) &1.5$\pm$0.7 (120) &2.9$\pm$1.4 (120) &-&130,131\\
\rule[0mm]{0mm}{4mm}
CI Cam = XTE J0421+560 &unknown&5$^{+3}_{-4}$ (114)&3.2$\pm$1.2 (121) &5$\pm$2 (124) &1.0&43,124,132\\
\rule[0mm]{0mm}{4mm}
LMC X--4 &NS &50$\pm$10 (3,4) &0.31$\pm$0.06 (122$^{\ast}$) &0.55$\pm$0.10 (122) &-&43,133\\
\rule[0mm]{0mm}{4mm}
A0535+26 = HDE 245770 &NS &2$^{+0.4}_{-0.7}$ (115) &2.3$\pm$0.5 (115)&11.8$\pm$1.5 (125) &1.0&125\\
\rule[0mm]{0mm}{4mm}
GX 301--2 = 4U 1223--62 &NS &5.3$\pm$0.1 (116) &5.9$\pm$0.6 (116) &20$\pm$10 (126) &-&116,126\\
\rule[0mm]{0mm}{4mm}
V4641 Sgr = SAX 1819.3--2525&BH &9.6$\pm$2.4 (5)&1.0$\pm$0.3 (5) &2.3$\pm$0.1 (127) &0.2&43,134,135\\
\rule[0mm]{0mm}{4mm}
KS 1947+300 = GRO J1948+32 &NS &10$\pm$2 (117) &3.38$\pm$0.16 (117) &34$\pm$30 (117) &1.0&43,117\\
\rule[0mm]{0mm}{4mm}
Cyg X--1 = HD 226868 &BH &2.1$\pm$0.1 (118) &2.95$\pm$0.21 (123) &6.21$\pm$0.22 (128)&1.0&43,136--138\\
\rule[0mm]{0mm}{4mm}
Cyg X--3 = V1521 Cyg &unknown&10.3$\pm$2.3 (119) &20$\pm$5 (119) &85$\pm$1 (129) &1.0&43, 139\\
\hline
\end{tabular}
\normalsize
\end{center}
Columns give:
(I) source names;
(II) BH = black hole, NS = neutron star;
(III) distance estimate;
(IV) interstellar reddening in $V$-band ($^{\ast}$$A_{\rm V}$ is estimated here from the relation $N_{\rm H} = 1.79 \times 10^{21} cm^{-2}A_{\rm V}$; \citealt{predet95});
(V) interstellar HI absorption column;
(VI) The maximum time separation, $\Delta$t, between the OIR and X-ray observations defined as quasi-simultaneous;
(VII) References for the quasi-simultaneous OIR and X-ray fluxes collected.
References for Tables 1, 2 \& 3:
(1) \cite*{stanga98};
(2) \cite{geliet03};
(3) \cite*{boydet00};
(4) \cite*{kova00};
(5) \cite*{jonket04};
(6) \cite{chatet03};
(7) \cite*{casaet04};
(8) \cite*{zdziet04};
(9) \cite*{dellet94};
(10) \cite*{cadoet04};
(11) \cite*{chapco04};
(12) \cite*{willet05};
(13) \cite*{hutcet03};
(14) \cite*{liuet01};
(15) \cite*{rittet03};
(16) \cite*{cowlet83};
(17) \cite*{geliet01};
(18) \cite{esinet00};
(19) \cite{oroset02};
(20) \cite*{hyneet98};
(21) \cite*{hyneet03};
(22) \cite{hyneet02a};
(23) \cite*{shraet97};
(24) \cite*{soriet01};
(25) \cite*{wilmet01};
(26) \cite*{wanget05};
(27) \cite*{konget02};
(28) \cite*{ebiset94};
(29) \cite*{kitaet90};
(30) \cite*{zdziet98};
(31) \cite*{tana93};
(32) \cite*{hyne05};
(33) \cite*{mcclet95};
(34) \cite{torret04};
(35) \cite*{oroset96};
(36) \cite*{oroset98};
(37) \cite{greeet01};
(38) \cite*{shahet01};
(39) \cite*{zuriet02};
(40) \cite*{casaet93};
(41) \cite{brocet04};
(42) \cite*{garcet01};
(43) \emph{RXTE} ASM;
(44) \cite{brocet01a};
(45) \cite{kuul98};
(46) \cite*{mcclet03};
(47) \cite*{kiziet05};
(48) \cite*{uemuet00};
(49) \cite*{hyneet05};
(50) \cite*{kinget96};
(51) \cite*{dellet98};
(52) \cite*{sutaet02};
(53) \cite*{brocet01b};
(54) \cite{buxtet04};
(55) \cite*{kaleet05};
(56) \cite*{hameet03};
(57) \cite{jainet01b};
(58) \cite*{market05};
(59) \cite*{torret05};
(60) \cite*{chatet02};
(61) \cite{corbet02};
(62) \cite{homaet05};
(63) \cite*{kuulet04};
(64) \cite*{israet04};
(65) \cite*{nagaet03};
(66) \cite*{tomset03};
(67) \cite*{haswet00};
(68) \cite*{fendpo00};
(69) \cite*{hyneet04};
(70) \cite*{zycket99};
(71) \cite{hanet92};
(72) \cite*{gallma05};
(73) \cite*{branet02};
(74) \cite*{rutlet02};
(75) \cite*{chevet89};
(76) \cite*{intzet01};
(77) \cite*{shawet05};
(78) \cite*{neleet04};
(79) \cite*{campet04};
(80) \cite*{vanset00};
(81) \cite*{wachet02};
(82) \cite*{steeet02};
(83) \cite*{campet02};
(84) \cite*{welset00};
(85) \cite*{torret02};
(86) \cite*{wanget01};
(87) \cite*{schu99};
(88) \cite*{hagget04};
(89) \cite*{blaiet84};
(90) \cite*{rutlet01};
(91) \cite*{grinli78};
(92) \cite*{campet05};
(93) \cite*{chevet99};
(94) \cite*{campet03};
(95) \cite*{shahet93};
(96) \cite*{burdet03};
(97) \cite*{torret06};
(98) \cite*{steeet04};
(99) \cite*{machet90};
(100) Russell et al. (in preparation);
(101) \cite*{kaluet80};
(102) \cite*{caniet80};
(103) \cite*{mcnaet03};
(104) \cite*{campst04};
(105) \cite*{homeet01};
(106) \cite*{charet80};
(107) \cite*{jainet99};
(108) \cite*{jainet00};
(109) \cite*{maitet03};
(110) \cite*{maitet04a};
(111) \cite*{maitet04b};
(112) \cite*{maitet05};
(113) \cite*{cole98};
(114) \cite*{miodet04};
(115) \cite*{steeet98};
(116) \cite*{kapeet95};
(117) \cite*{neguet03};
(118) \cite*{masset95};
(119) \cite*{vanket96};
(120) \cite*{lequet92};
(121) \cite*{hyneet02b};
(122) \cite*{naiket03};
(123) \cite*{wuet82};
(124) \cite*{orlaet00};
(125) \cite*{orlaet04};
(126) \cite*{mukhet04};
(127) \cite*{dicket90};
(128) \cite*{schuet02};
(129) \cite*{staret03};
(130) \cite*{claret78};
(131) \cite*{cowlet04};
(132) \cite*{claret00};
(133) \cite*{heemet89};
(134) \cite*{katoet99};
(135) \cite*{buxtet05b};
(136) \cite*{bochet98};
(137) \cite*{brocet99b};
(138) \cite*{brocet99a};
(139) \cite*{kochet02}
\end{table*}

The X-ray unabsorbed 2-10 keV flux was calculated for all X-ray data\footnote{This energy range was adopted to be consistent with that used for the radio--X-ray correlations of \citeauthor{gallet03} (2003; 2--11 keV) and \citeauthor{miglfe06} (2005; 2--10 keV).}. We made note of the X-ray state of each source on each observation, as defined by the analysis by authors in the literature. We were unable to apply strict definitions to the spectral states due to the differing nature (e.g. X-ray energy ranges and variability analysis) of each data set. We have therefore used the judgement of the authors to determine the spectral states; a method likely to be much more accurate than any conditions that could be imposed by us. We assume a power-law with a spectral index $\alpha=-$0.6 (photon index $\Gamma=1.6$) where $F_{\nu}\propto \nu^{\alpha}$, when the source is in the hard state, and a blackbody at a temperature of 1 keV for soft state data. These models for the X-ray spectrum are the same as those adopted by \cite{gallet03}, and altering the values in the models to other reasonable approximations does not significantly change the estimated X-ray luminosities. For GRS 1915+105, we use data from the radio-bright plateau state, which is approximately analogous to the hard state \citep[e.g.][]{fendpo00}. The \emph{NASA} tool \emph{Web-PIMMS} was used to convert from instrument X-ray counts per unit time (e.g. day-averaged $RXTE$ ASM counts s$^{-1}$). \cite{brocet04} also provide a table of approximate instrument counts--flux conversion factors used in this work.

For the OIR luminosities, data was collected from the optical $B$-band at 440 $nm$ to the near-infrared $K$-band at 2220 $nm$. OIR absorbed fluxes were de-reddened using the best-known value of the extinction $A_{\rm V}$ to each source and the dependence of extinction with wavelength given by \cite*{cardet89}. For OIR data at fluxes low enough for the companion star to significantly contribute (i.e. $quiescence$ for most low-mass XBs), the data was discarded if the fractional contribution of the companion had not been estimated in the OIR band of the data. The fractional contribution of the companion (column 7 of Table 1) was subtracted from the low-flux OIR data in order to acquire the flux from all other emission processes in the system. This contribution is not well constrained in many cases due to the uncertain spectral type of the companion star \citep*[e.g.][]{haswet02}. We have therefore propagated the errors associated with this into the errors of the OIR luminosities for all quiescent data. In sources for which the companion is comparatively bright \citep*[e.g. GRO J1655--40;][]{hyneet98}, its contribution has been subtracted in outburst in addition to quiescence.

The intrinsic (de-reddened) OIR and X-ray luminosities were then calculated given the best-known estimate of the distance to each source (Table 1). We adopted the approximation $L_{\rm OIR}\approx \nu F_{\nu,OIR}$ to estimate the OIR luminosity (we are approximating the spectral range of each filter to the central wavelength of its waveband). The errors associated with the luminosities are propagated from the errors quoted in the original data. Where no errors are quoted, we apply a conservative error of 30\%. Errors associated with estimates of the distance, extinction $A_{\rm V}$ and NI absorption column $N_{\rm H}$ were sought (Table 1). Where these limits are not directly quoted in the reference, we used the most conservative estimates implied from the text. These errors were not propagated into the errors associated with the luminosities we derive because the resulting plots would be dominated by error bars, however we also show one error bar for each data set, representing the average total systematic 1$\sigma$ errors associated with each luminosity data point.

In addition, we searched the literature for quasi-simultaneous OIR and X-ray fluxes from a number of NSXBs in a hard X-ray state \citep[mostly atoll sources in the `island state', defined by the X-ray colour--colour diagram; see e.g.][]{hasiva89,miglfe06}, and HMXBs in both the hard and soft X-ray state. The same methodology was adopted in calculating the intrinsic OIR and X-ray luminosities of the NSXBs and the HMXBs. Tables 2 and 3 list the properties of the NSXBs and HMXBs, respectively. A literature search for OIR Spectral Energy Distributions (SEDs) of BHXBs was also conducted in order to shed further light on the nature of the emission. Fluxes were used when two or more OIR wavebands were quasi-simultaneous. No X-ray fluxes were required for the SEDs, however the X-ray state of the source on the date was noted. Where the companion star significantly contributes to the emission, the estimated wavelength-dependent contribution of the companion was subtracted (adopting the quoted contributions given in the papers from which the data was acquired).

\subsection{Results}

Quasi-simultaneous OIR and X-ray luminosities are plotted of 15 BHXBs in the hard state (Fig. 1$a$), 9 BHXBs in the soft state (Fig. 1$b$), 8 NSXBs in the hard state (Fig. 2$a$) and 9 HMXBs (Fig. 2$b$). We have classed LMC X--3 as a low-mass XB (BHXB) because \cite{brocet01a} found from 6 years of observations on this source that its optical emission is dominated by long-term variations rather than the bright companion star in the system, due to its persistent nature (and we are using data from their paper).

For BHXBs in the hard state, a strong $L_{\rm OIR}$--$L_{\rm X}$ correlation exists over 8 orders of magnitude in X-ray luminosity. The data from all 15 individual sources lie close to this correlation but deviations in the slopes and normalisations of individual sources are present. The slope of the global correlation is $\beta$=0.61$\pm$0.02, where $L_{OIR} \propto L_{\rm X}^{\beta}$ (we do not take into account the $L_{\rm OIR}$ and $L_{\rm X}$ error bars in calculating $\beta$). To calculate correlations we use the package \emph{gnuplot} and apply equal weighting to each data point. The relations will be biased towards sources with more data points, but we argue that all data points are equally important because we have used a maximum of one data point per day per waveband per source, and the sources tend to vary on timescales $\geq$ days.

In the soft state, the optical data ($BVRI$-bands) lie close to the hard state correlation for BHXBs (Fig. 1$b$), and the NIR data ($JHK$-bands) lie below the correlation. A correlation is also found for the NSXBs (Fig. 2$a$) over 7 orders of magnitude in X-ray luminosity. Its slope, $L_{\rm OIR} \propto L_{\rm X}^{0.63\pm0.04}$, is remarkably similar to the OIR--X-ray correlation of the BHXBs, but with a lower normalisation. Taking the two correlations, we find that at a given X-ray luminosity (2--10 keV), the OIR luminosity of a BHXB in the hard state is on average 19.4 times larger (1.29 dex) than that of a NSXB in the hard state in the X-ray range of the data.

If we neglect the data from sources in quiescence ($L_{\rm X}<10^{33.5}$ erg s$^{-1}$), the hard state fits are $L_{\rm OIR} = 10^{13.2\pm 0.8} L_{\rm X}^{0.60\pm0.02}$ for BHXBs and $L_{\rm OIR} = 10^{9.7\pm 1.8} L_{\rm X}^{0.66\pm0.05}$ for NSXBs. These are very similar to those with quiescent data included. The similarity of the fits strengthens the case for the quiescent state being an extension to the hard state. For the analysis in the following Sections, we use the fits with quiescent data included, as they are similar enough either way. In addition, it is also interesting that the fit to the BHXB quiescent data alone has a slope $L_{OIR} \propto L_{\rm X}^{0.67\pm 0.14}$ (there are only 4 quiescent NSXB data points).

From Fig. 2$b$ it is clear that the OIR luminosity of HMXBs is typically orders of magnitude larger than that of LMXBs, and does not appear to correlate with $L_{\rm X}$. This is consistent with the high-mass companion star dominating the OIR emission. The large range in OIR luminosities between sources is likely to be due to the differing masses and spectral types of the companions. Since this interpretation agrees with the evidence in the literature, we will not discuss the HMXBs in further detail (although they are mentioned in Section 3.5).

There are clear advantages and disadvantages of these OIR--X-ray compilations compared to the $L_{\rm radio}$--$L_{\rm X}$ approach of \cite{gallet03}. The hard state OIR--X-ray correlation in BHXBs includes data from many BHXBs in low-luminosity \emph{quiescent} states (7 sources with $10^{30}<L_{\rm X}<10^{33.5}$ erg s$^{-1}$), which was not possible for the radio--X-ray correlation due to radio detector limits \citep[but see][]{gallet06}. The BHXB sample also includes sources in the LMC and M31, which are too distant (and hence faint) to observe at radio wavelengths. However, unlike the radio--X-ray comparison, HMXBs cannot be included in these OIR--X-ray plots as the OIR emission is dominated by the companion. In addition, two sources situated in the galactic plane (1E 1740.7--2942 and GRS 1758--258) were included in the radio--X-ray correlation but are not observable at OIR wavelengths due to the high levels of extinction towards the sources.

\begin{figure*}
\includegraphics[height=22cm,width=17cm,angle=0]{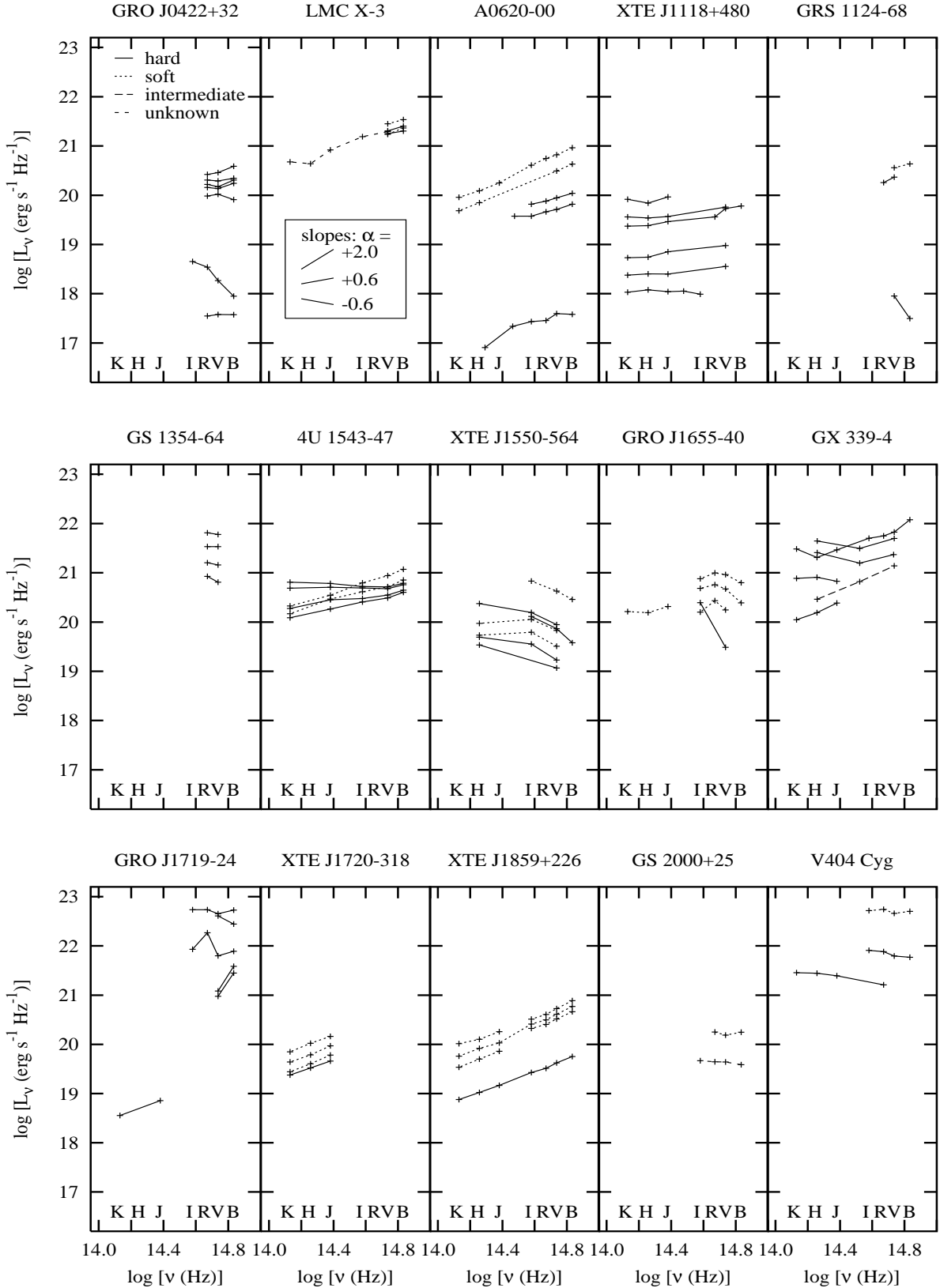}
\caption{Spectral Energy Distributions (SEDs) of 15 BHXBs. The key in the top left panel corresponds to the X-ray state of the source on the date of observation. A key to the slope of the continuum, i.e. the spectral index $\alpha$ (where $L_{\nu} \propto \nu^{\alpha}$), is indicated in the second panel on the top row. The dates of observations and references for the data are listed in Table 4.
}
\end{figure*}

\begin{table*}
\begin{center}
\caption{Dates and references for the data in Fig. 3.}
\begin{tabular}{llllllll}
\hline
Source&X-ray&Dates&X-ray&Dates&X-ray&Dates&References\\
      &state&(MJD)&state&(MJD)&state&(MJD)&\\
\hline
GRO J0422+32 &hard&\multicolumn{3}{l}{48866, 48870, 49006,7, 49039, 49209, 49590}& & &1 - 4\\
LMC X--3 &hard&50151, 50683 &soft&50324, 51038 &unknown&46804&5 - 7\\
A0620--00 &hard&42792, 43097-100, 42859 &soft&42650-2, 42703 & & &8 - 11\\
XTE J1118+480 &hard&\multicolumn{3}{l}{51652, 51975, 53385-6, 53393, 53409, 53412}&& &12, 13$^2$\\
GRS 1124--68 &hard&48453, 48622 &soft&48367-71 & & &14 - 16\\
GS 1354--64 &hard&50778, 50782, 50851, 50888& & & & &17\\
4U 1543--47 &hard&52486, 52490, 52495, 52501&soft&52454, 52469 & & &18\\
XTE J1550--564&hard&51260, 51630, 51652, 51717&soft&51210, 51660, 51682& & &19, 20\\
GRO J1655--40 &hard&53422 &soft&\multicolumn{3}{l}{50217, 50254, 50286, 50648}&21 - 23\\
GX 339--4 &hard&\multicolumn{3}{l}{44748-59, 49539, 50648, 52368, 52382}&intermediate&52406&23 - 25\\
GRO J1719--24 &hard&\multicolumn{3}{l}{49096, 49101, 49108, 49266, 49280, 49539}&& &23, 26 - 29\\
XTE J1720--318&hard&52782 &soft&52659, 52685, 52713& & &30\\
XTE J1859+226 &hard&51605-8 &soft&\multicolumn{3}{l}{51465,9, 51478, 51488, 51501}&31, 32\\
GS 2000+25$^1$&soft&47360 &unknown&47482 & & &33, 34\\
V404 Cyg &hard&47684, 47718-27 &soft&47678 & & &35 - 38\\
\hline
\end{tabular}
\normalsize
\end{center}
$^1$ This source (alternative name: QZ Vul) is not tabulated in Table 1; its distance and reddening are 2.7$\pm$0.7 kpc and $Av\sim$ 3.5, respectively \citep{jonket04}. $^2$This previously unpublished data was obtained with the Liverpool Telescope and the United Kingdom Infrared Telescope. See Brocksopp et al. (in preparation) for the data reduction recipe used. References:
(1) \cite{bartet94};
(2) \cite*{goraet96};
(3) \cite*{castet97};
(4) \cite*{hyneha99};
(5) \cite{brocet01a};
(6) \cite*{trevet87};
(7) \cite*{trevet88};
(8) \cite*{robeet76};
(9) \cite*{okeet77};
(10) \cite{oke77};
(11) \cite*{kleiet76};
(12) \cite{chatet03};
(13) this paper (see Brocksopp et al. in preparation);
(14) \cite{kinget96};
(15) \cite{dellet98};
(16) \cite{bail92};
(17) \cite{brocet01b};
(18) \cite{buxtet04};
(19) \cite{jainet01a};
(20) \cite{jainet01b};
(21) \cite*{buxtet05a};
(22) \cite{hyneet98};
(23) \cite{chatet02};
(24) \cite{corbet02};
(25) \cite{homaet05};
(26) \cite{sekiwy93};
(27) \cite{allejo93};
(28) \cite{allegi93};
(29) \cite{brocet04};
(30) \cite{nagaet03};
(31) \cite{haswet00};
(32) \cite{hyneet02a};
(33) \cite{charet81};
(34) \cite*{chevil90};
(35) \cite*{szkoet89};
(36) \cite*{gehret89};
(37) \cite{casaet91};
(38) \cite{hanet92}

\end{table*}

In Section 3 we attempt to interpret the relations found between OIR and X-ray luminosities in terms of the most likely dominant emission processes. OIR SEDs were collected from 15 BHXBs in a range of luminosities and X-ray states and are presented in Fig. 3 and the dates of observations and references are given in Table 4. The SEDs are interpreted in Section 3.3. In Sections 3.4 and 3.5 we discuss additional patterns, applications and implications of the empirical relations. The results and interpretations are summarised in Section 4.

\begin{figure*}
\includegraphics[height=15.5cm,angle=270]{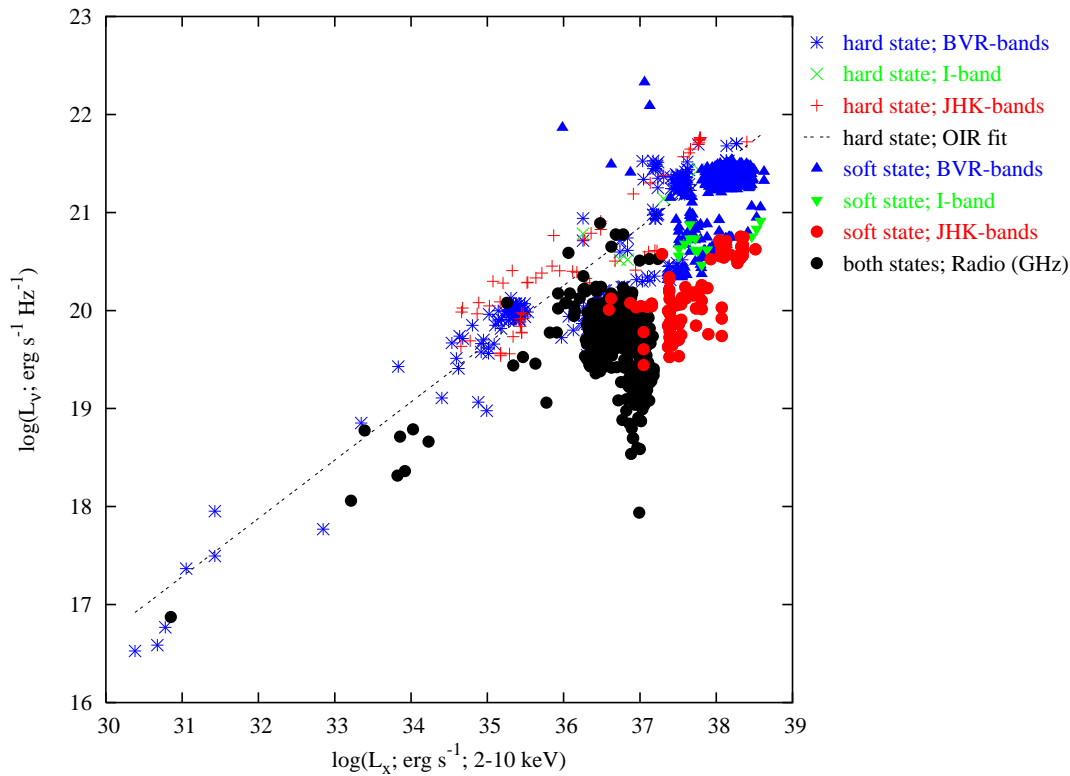}
\caption{X-ray 2--10 keV luminosity versus OIR and radio monochromatic luminosities for BHXBs.}
\end{figure*}

\section{Interpretation \& Discussion}

\subsection{Jet Suppression in the Soft State}

The significant drop in some of the OIR data in the soft state for BHXBs compared to the hard state (Fig. 1$b$) is not due to changes in the X-ray luminosity during state transition because although the X-ray spectrum significantly changes during transition, the bolometric (and 2--10 keV) X-ray luminosity does not \citep[e.g.][]{zhanet97,kordet06}. The jet component at radio wavelengths is known to decrease/increase in transition to/from the soft X-ray state, due to the quenching of the jets in the soft state \citep[e.g.][]{gallet03}. In Fig. 4 we split the OIR data from BHXBs into three wavebands: $BVR$-bands, $I$-band and $JHK$-bands, and plot the monochromatic OIR luminosity ($L_{\nu}$; i.e. flux density scaled for distance) against $L_{\rm X}$, overplotted with the radio data $L_{\rm \nu ,radio}$--$L_{\rm X}$ for BHXBs from \citeauthor{gallet03} (2003; 2006). We see that the normalisations in $L_{\nu}$ (as well as the power-law slopes) for the radio--X-ray and OIR--X-ray hard state correlations are similar to one another for BHXBs: at a given $L_{\rm X}$, the radio and OIR monochromatic luminosities are $\sim$ equal, implying a flat spectrum from radio to OIR wavelengths for all BHXBs in the hard state. We return to the hard state interpretation in Section 3.2.

Fig. 4 also shows a clear suppression of all the $JHK$ data, and a little, if any of the $BVR$ in the soft state. The $I$-band data appear to sit in the centre of the two groups. All 9 BHXBs for which we have soft state data are consistent with this behaviour, suggesting it is ubiquitous in BHXBs. We interpret this as the NIR wavebands being quenched as the jet is switched off, as is observed at radio wavelengths. The NIR appears to be quenched at a higher X-ray luminosity than the radio, but this effect may be just due to an upper $L_{\rm X}$ limit adopted by \cite{gallet03} when compiling their radio--X-ray data.

The optical wavebands in the soft state lie close to the OIR hard state correlation. Most of the optical data lies below the centre of the hard state correlation, with the exception of V404 Cyg, whose optical luminosity is enhanced in the soft state. There is still a debate as to whether V404 Cyg entered the soft state, and the data here were taken close to the supposed state transition. The $I$-band appears to be the ``pivot'' point, as already shown by \cite{corbet02} and \cite{homaet05} to be where the continuum of the optically thin jet meets that of (possibly) the disc. The NIR quenching in the soft state implies that this waveband is dominated by the jets in luminous hard states, just before/after transition to/from the soft state. The optical data are not quenched, suggesting a different process is dominating at these wavelengths.

An alternative interpretation could be that the disc dominates the OIR in both the hard and soft states, but changes temperature during state transition, shifting the blackbody from (e.g.) the OIR in the hard state, to the optical--UV in the soft state. This would have the effect of reducing the NIR in the soft state but maintaining the optical, as is observed. We argue that this is not the case because there is evidence in many BHXBs for two spectral components of OIR emission, the redder of which is quenched in the soft state \citep[Section 3.3;][]{jainet01b,buxtet04,homaet05}.

The jets should contribute negligible OIR flux in the soft state, so we can estimate the OIR contribution of the jets at high luminosity in the hard state from the level of soft state quenching. The mean offset of the OIR soft state data from the hard state fit is 0.30$\pm$0.32 dex, 0.71$\pm$0.21 dex and 1.10$\pm$0.26 dex in $L_{\rm\nu,OIR}$ for the $BVR$, $I$ and $JHK$-bands, respectively. This corresponds to a respective fractional jet component of 50$^{+26}_{-50}$, 81$^{+7}_{-13}$ and 92$\pm$5 percent in the $BVR$, $I$ and $JHK$-bands. It is clear that the spectrum of the jet extends, with spectral index $\alpha\sim 0$, from the radio regime to the NIR in BHXBs. The position of the turnover from optically thick to optically thin emission must be close to the NIR waveband for the radio--NIR spectrum to appear flat \citep[unless the optically thick spectrum is highly inverted, which is not seen in the radio spectrum most of the time; see e.g.][]{fend01}. The OIR spectrum in the hard state is flatter than in the soft state, as is expected if the jet component is present in the former and not in the latter; this is explored further in Section 3.3.

\subsection{Modelling Disc and Jet Contributions in the Hard State}

\begin{figure*}
\centering
\includegraphics[height=15.5cm,angle=270]{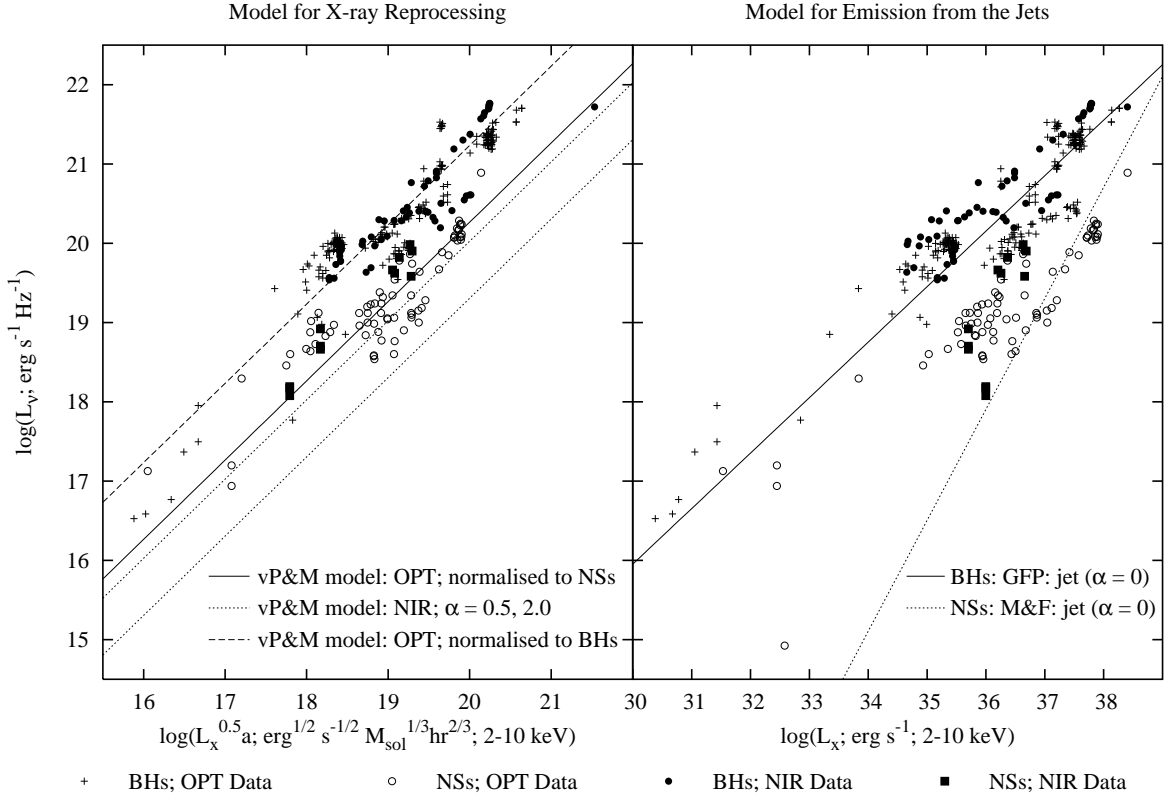}
\caption{BHXBs and NSXBs in the hard X-ray state, with models for X-ray reprocessing (left) and jet (right) components. The X-ray reprocessing models are derived from vP\&M \citep{vanpet94} and the jet models assume a flat spectrum from radio to OIR and use the relations from GFP \citep{gallet03} for BHXBs and M\&F \citep{miglfe06} for NSXBs.}
\end{figure*}

We now attempt to interpret the empirical hard state correlations in terms the three most cited OIR emission processes: X-ray reprocessing in the disc, the viscously heated disc and jet emission. We adopt the theoretical model between optical and X-ray luminosities of \cite{vanpet94} for X-ray reprocessing (Section 1.1). \citeauthor{vanpet94} normalised their relation to a sample of systems, some of which are BHXBs and some NSXBs. Here, we normalise the relation to our sample of NSXBs (just the optical $BVRI$-bands), which is much larger than the sample of \citeauthor{vanpet94}. There is more concrete evidence for NSXBs to be dominated by X-ray reprocessing in the optical regime than in BHXBs (see Section 1). We also take into account a dependency of the relation on system mass that was neglected by \citeauthor{vanpet94}, but is needed here to compare between NSXBs and BHXBs. From Kepler's third law and equation 5 of \citeauthor{vanpet94} (and the discussion that follows), we have:
\begin{eqnarray*}
  L_{\rm OPT}\propto L_{\rm X}^{1/2}a \propto L_{\rm X}^{1/2}(M_{\rm co}+M_{\rm cs})^{1/3}P^{2/3}
\end{eqnarray*}

In the left panel of Fig. 5 we plot the monochromatic luminosity (optical and NIR) versus $L_{\rm X}^{1/2}a$ for each BHXB and NSXB data point in the hard state (adopting the values of $P$, $M_{\rm co}$ and $M_{\rm cs}$ for each source from Tables 1 and 2). The solid line represents the power-law fit to the optical NSXB data, fixing the slope at unity ($L_{\rm OPT}\propto L_{\rm X}^{1/2}a$). We expect a lower normalisation for the NIR data due to the shape of the OIR spectrum, which may have a spectral index $0.5\leq\alpha\leq2.0$ \citep[a conservative range based on theoretical and empirical results of X-ray reprocessing; see e.g.][]{hyne05}. The upper and lower dotted lines indicate the expected correlations of the NIR data (approximating the optical to the $V$-band centred at $550nm$ and the NIR to the $H$-band at $1660 nm$) if $\alpha =0.5$ and $2.0$, respectively. The dashed line shows the fit to the optical BHXB data (fixing the slope at $L_{\rm OPT}\propto L_{\rm X}^{1/2}a$), which appears to be elevated by $\sim 1$ order of magnitude in $L_{\rm \nu,OIR}$ compared to the optical NSXB data. Similarly, the NIR data of both BHXBs and NSXBs lie above the expected correlation for emission from X-ray reprocessing. The implications of these fits are discussed below.

For the jets, we take the models described in Section 1.1 where the spectrum is flat from radio to OIR, and normalise them using the empirical $L_{\rm radio}$--$L_{\rm X}$ relations found by \cite{gallet03} and \cite{miglfe06} for BHXBs and NSXBs, respectively. In the right panel of Fig. 5 we plot $L_{\rm \nu ,OIR}$ versus $L_{\rm X}$ for hard state BHXBs and NSXBs. The same relation is expected between optical and NIR jet emission when plotting monochromatic luminosity because we are assuming the jet spectrum is flat; $\alpha \sim 0$. We find that the jet models can approximately describe the optical and NIR data of the BHXBs. The NIR data from NSXBs lie above the expected jet correlation but possess a similar slope, whereas the slope of the optical NSXB data is very different to that predicted from jet emission.

The models for OIR emission from a viscously heated disc are described as follows. For the simplest viscously heated steady-state disc, there are two limiting regimes \citep{franet02}. For $h\nu \ll kT$, we expect $L_{\rm OIR}\propto$ \.m$^{1/4}$. This is simply the Rayleigh-Jeans limit and will only apply well into the IR. For $h\nu \gg kT$, the relationship is steeper, $L_{\rm OIR}\propto$ \.m$^{2/3}$. For typical disc edge temperatures of 8,000--12,000\,K, expected power-law slopes ($L_{\rm OIR}\propto$ \.m$^{\gamma}$), are calculated to vary from $\gamma\sim0.3$ in the $K$-band, to $\gamma\sim0.5$ in $V$, and $\gamma\sim0.6$ in the UV. Using the calculations linking $L_{\rm X}$ and \.m in Section 1.1, this corresponds to expected correlations of the form $L_{\rm OIR}\propto L_{\rm X}^{\beta}$, where 0.15$\le\beta\le$0.25 for BHXBs and 0.30$\le\beta\le$0.50 for NSXBs.

\begin{table*}
\caption{Parameters for the hard state models: theory versus observations.}
\begin{tabular}{lrcrrcrrc}
\hline
\multicolumn{1}{c}{Sample}&\multicolumn{3}{c}{------------ X-ray reprocessing model ------------}&\multicolumn{3}{c}{----------------- Jet model -----------------}&\multicolumn{2}{c}{-- Viscous disc model --}\\
      &\multicolumn{1}{c}{Model}&\multicolumn{1}{c}{$\mid \beta_{\rm data}$-$\beta_{\rm model}\mid$}&\multicolumn{1}{c}{$\frac{n_{\rm data}}{n_{\rm model}}^\ast$}&\multicolumn{1}{c}{Model}&\multicolumn{1}{c}{$\mid \beta_{\rm data}$-$\beta_{\rm model}\mid$}&\multicolumn{1}{c}{$\frac{n_{\rm data}}{n_{\rm model}}$}&\multicolumn{1}{c}{Model}&\multicolumn{1}{c}{$\mid \beta_{\rm data}$-$\beta_{\rm model}\mid$}\\
\hline
\vspace{2mm}
BHs: $L_{\rm \nu,OPT}$&$n L_{\rm X}^{0.5}a$&0.05$\pm$0.03&9.3$\pm$0.4&$L_{\rm X}^{0.7}$&0.11$\pm$0.02&1.05$\pm$0.07&$L_{\rm X}^{0.25}$&0.34$\pm$0.02\\
\vspace{2mm}
BHs: $L_{\rm \nu,NIR}$&$(\frac{\nu_{NIR}}{\nu_{OPT}})^\alpha n L_{\rm X}^{0.5}a$&0.06$\pm$0.03&15.5--81.3$^\dagger$ &$L_{\rm X}^{0.7}$&0.09$\pm$0.04&1.78$\pm$0.16&$L_{\rm X}^{0.17}$&0.44$\pm$0.04\\
\vspace{2mm}
NSs: $L_{\rm \nu,OPT}$&$n L_{\rm X}^{0.5}a$&0.09$\pm$0.02&1.0$^\ast$ &$L_{\rm X}^{1.4}$&0.81$\pm$0.03&6.03$\pm$1.94&$L_{\rm X}^{0.50}$&0.09$\pm$0.03\\
NSs: $L_{\rm \nu,NIR}$&$(\frac{\nu_{NIR}}{\nu_{OPT}})^\alpha n L_{\rm X}^{0.5}a$&0.05$\pm$0.03&3.2--16.6$^\dagger$ &$L_{\rm X}^{1.4}$&0.09$\pm$0.41&9.55$\pm$3.08&$L_{\rm X}^{0.30}$&1.19$\pm$0.41\\
\hline
\end{tabular}
\normalsize
For $\mid \beta_{data}$-$\beta_{model}\mid$, $\beta$ and $n$ are free parameters; for $\frac{n_{data}}{n_{model}}$, $\beta$ is fixed at the value of the model and $n$ is a free parameter; $^\ast$ $n$ is defined by the fit to the optical NSXB data (see text); $^\dagger$ the range corresponds to an OIR spectral index $0.5\leq\alpha\leq 2.0$.
\end{table*}

A summary of the results of fitting these models to the observed data is provided in Table 5. It is clear that for BHXBs, the slopes $\beta$ of the observed relations can be explained by the X-ray reprocessing model or the jet model, but not by the viscous disc model (Homan et al. 2005 also ruled out a viscous disc origin to the OIR emission in the hard state of GX 339--4 because the mass accretion rate inferred from the luminosity is much higher than expected). This is also true for the NIR data of the NSXB sample, however the slope of the NSXB optical data cannot be explained by the jet model and are accurately described by both the reprocessing and viscous models. At high luminosities in the NSXB sample, no OIR data points lie far below the jet relation, suggesting that this process may indeed play a significant OIR role here \citep[as is seen in a few NSXBs, see][ and references therein]{miglet06}.

The normalisation $n$ of the BHXB data is closer to the jet model than the reprocessing model. Although this is consistent with the constraints derived from Section 3.1 (the jets are contributing $\sim90$ percent of the NIR luminosity here), an optical excess of $\sim$1 order of magnitude from the reprocessing model is not expected. The excess is unlikely to be fully explained by the jets because of the lack of optical quenching in the soft state (Section 3.1). In addition, the optical spectrum of most BHXBs is inconsistent with jet emission dominating (Section 3.3). Instead we suggest that OIR emission from reprocessing is enhanced for BHXBs at a given $L_{\rm X}$ due to the localisation of the source of X-rays. For example, the X-ray emitting region in BHXBs may have a larger scale height than in NSXBs and will therefore illuminate the disc more readily \citep[e.g.][]{minifa04}. This may account for the high value of $n$ for the optical BHXB data, but still struggles to explain the even higher $n$ for the NIR BHXB data.

We note that there are deviations of individual sources from the correlations which may be caused by distance and reddening errors, or by differing emission process contributions due to the range of orbital separations or slight differences in the slope of the radio--OIR jet spectrum between sources (or other system parameters not considered). For example, XTE J1118+480 has a small disc and is known to produce significant optical jet emission \citep*[e.g.][]{malzet04,hyneet06} whereas V404 Cyg possesses a large disc and is dominated by X-ray reprocessing in the disc \citep[e.g.][]{wagnet91}. Other OIR emission mechanisms that could contribute, for example disc OIR emission due to magnetic reconnection, have not been modelled here and could also contribute to the scatter in Fig. 5. These processes cannot be ruled out, but are unlikely to easily explain the observed correlations.

\begin{figure}
\centering
\includegraphics[width=6cm,angle=270]{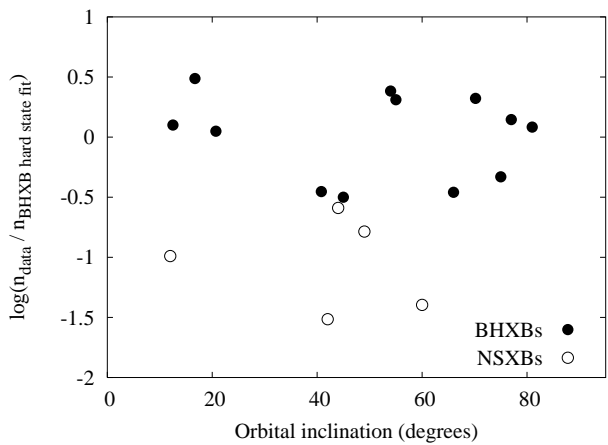}
\caption{Mean normalisation $n$ of each source, fixing the slope at $L_{\rm OIR}\propto L_{\rm X}^{0.6}$, versus orbital inclination.}
\end{figure}

In addition, the orbital inclination may affect the level of OIR emission, in particular from X-ray reprocessing in the disc. To explore this, we have plotted in Fig. 6 the average normalisation $n$ of the data for each source against the known estimated orbital inclination, fixing the slope at $L_{\rm OIR}\propto L_{\rm X}^{0.6}$ for each source. The best estimate of the inclination of each system (where known) was obtained from \cite{rittet03}, \cite{trevet88}, \cite*{fomaet01}, \cite{wanget01} and \cite{falaet05}. We find no evidence for there being a direct relation between $L_{\rm OIR}$ and the orbital inclination $i$, suggesting the effect of inclination is subtle.

\subsection{The Spectral Energy Distributions}

Although no clear patterns are visible from the SEDs in Fig. 3 on first inspection, closer analysis reveals support for many of the conclusions made so far. Optically thin synchrotron emission is expected to produce an OIR spectral index $\alpha<$ 0, and this is seen in part of the SEDs of 10 out of 14 BHXBs in the hard state. In the soft state only 3 out of 10 BHXBs are observed to have $\alpha<$ 0, suggesting a synchrotron component playing a larger role in the hard state than in the soft \citep[$\alpha$ in some sources could be dominated by uncertainties in $A_{\rm V}$; e.g. the SEDs of XTE J1550--564 are red even in the soft state, which is likely due to an underestimated extinction; see][]{jonket04}. In comparison, $\alpha>$ 0 is seen in the SEDs of 10/14 BHXBs in the hard state and 10/10 in the soft state. These spectra are likely to have a thermal origin and agree with recent analysis of optical/UV SEDs of 6 BHXBs \citep{hyne05}. The SEDs generally appear redder in the hard state than in the soft, as is observed by the NIR suppression in the soft state (Section 3.1).

A clear suppression of the NIR, and not the optical bands in the soft state is visible in the SEDs of GX 339--4, XTE J1550--564 and 4U 1543--47 \citep{homaet05,jainet01b,buxtet04}. In contrast, the value of $\alpha$ in XTE J1720--318 and XTE J1859+226 appear not to change between the hard and soft states. The SEDs show no substantial evidence for the turnover in the jet spectrum from optically thick ($\alpha\sim$ 0) to optically thin ($\alpha\sim-$0.6) synchrotron emission, as we may expect to see in the hard state \citep[except in GX 339--4; see][]{corbet02}. This is consistent with the turnover lying close to the NIR (Section 3.1), possibly just redward of $K$-band. In some systems, $\alpha$ is more negative at low luminosities. This effect could be the result of (a) a cooler accretion disc, or (b) the jets contributing more than the disc at low luminosities. Since the former process cannot explain the steeply negative SEDs of a few BHXBs at low luminosities, we suspect both processes may play a role.

Finally, the hard state NIR excess seen in GX 339--4 (\citealt{corbet02,homaet05}; interpreted as where the optically thin jet spectrum meets the blue thermal spectrum) is here also seen in LMC X--3 (tentatively), 4U 1543--47, XTE J1550--564 and V404 Cyg. In these sources, we interpret the NIR excess as originating in the optically thin part of the jet spectrum. It is interesting to note that OIR SEDs of a number of NSXBs show an IR excess that cannot be explained by thermal emission, and is likely to originate in the jets \citep[][ and references therein]{miglet06}.

\subsection{The Rise and Fall of an Outburst}

During a transient outburst typical of BHXBs, the source will either remain in the hard state for the entire outburst \citep[e.g. XTE J1118+480; GRO J1719--24; see also][]{brocet04} or transit into the soft (or intermediate or very high) state, before returning to the hard state and declining in luminosity. A hysteresis effect has been identified in BHXB transients, and at least one NSXB transient: the hard to soft state transition always occurs at a higher X-ray luminosity than the transition back to the hard state (\citealt*{maccco03,maccet03,yuet03}; \citealt{fendet04,homabe05}).

Here, we present a prediction of the OIR behaviour of an outburst that results from this hysteresis effect and the models in Section 3.2. The following sequence of events should occur for a BHXB outburst that enters the soft state:

\begin{itemize}
\item $L_{\rm OIR}$ and $L_{\rm X}$ increase in the hard state rise of the outburst, with jets and reprocessing contributing to the OIR.
\item The source enters the soft state, quenching the jet component. $L_{\rm OIR}$ drops but $L_{\rm X}$ is maintained.
\item $L_{\rm OIR}$ and $L_{\rm X}$ decrease before transition back to the hard state, when the jet component returns.
\item $L_{\rm OIR}$ and $L_{\rm X}$ continue to decline, with jets and X-ray reprocessing contributing towards $L_{\rm OIR}$.
\end{itemize}

\begin{figure}
\centering
\includegraphics[width=8.48cm,angle=0]{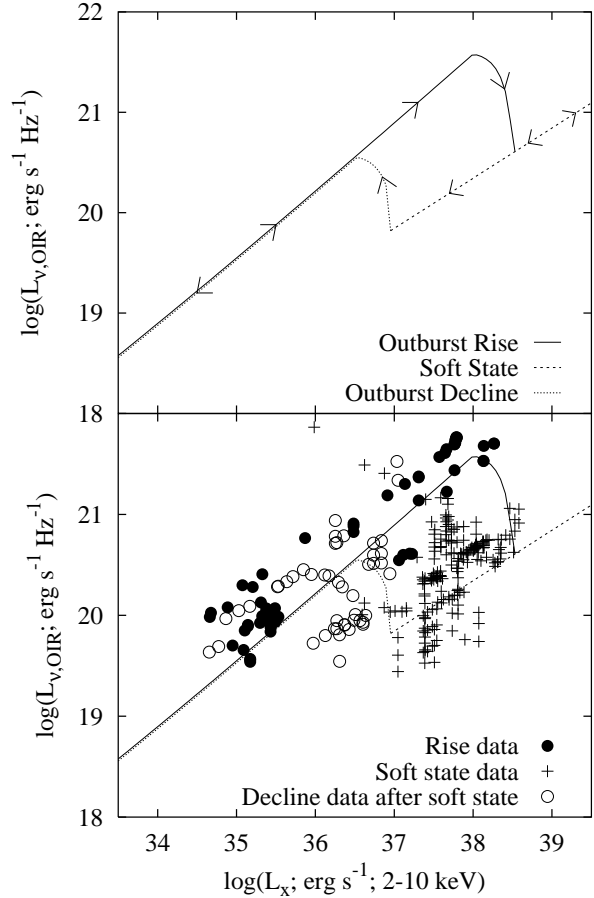}
\caption{Top panel: Schematic of the expected $L_{\rm OIR}$--$L_{\rm X}$ behaviour for a BHXB outburst that enters the soft state. Lower panel: As the top panel but with OIR data from the rise, soft state and decline (for sources that entered the soft state) stages of outbursts.}
\end{figure}

This sequence is illustrated in the top panel of Fig. 7. In this schematic, we fix the hard-to-soft state transition at $L_{\rm X}\sim 10^{38}$ erg $s^{-1}$ and the soft-to-hard, at $L_{\rm X}\sim 10^{37}$ erg $s^{-1}$. To test our hysteresis prediction, we split the hard state BHXB data into data from the rise of outbursts and data from the declines. We define rise and decline as before and after the peak X-ray (2--10 keV) luminosity of the outburst, respectively. We do not include data on the decline for sources that remained in the hard state throughout the outburst, as this is not what our prediction is testing. In the lower panel of Fig. 7 we plot these and the soft state data. Data from persistent sources (LMC X--3 and GRS 1915+105) and from sources in or near quiescence are not included. We find that the prediction is consistent with the data, with some inevitable scatter from errors as described in Section 3.2. The loop is larger in the NIR data, as is expected because the jets contribute more in the NIR than in the optical regime. Other reasons for any deviations from the expected models are also discussed in Section 3.2.

\subsection{Applications of the Correlations}

The existence of the OIR--X-ray correlations leads to a number of intriguing tools and uses for quasi-simultaneous multi-wavelength data.
\newline\newline
\textbf{The Mass Accretion Rate:}\newline
In Section 1.1 we show how $L_{\rm X}$ and \.m are thought to be linked for BHXBs and NSXBs. Here, we can use the empirical hard state $L_{\rm OIR}$--$L_{\rm X}$ correlations to link the OIR luminosity to the mass accretion rate:

\smallskip BHXBs: $L_{\rm OIR}\propto L_{\rm X}^{0.6}\propto$ \.m$^{1.2}$

\smallskip NSXBs: $L_{\rm OIR}\propto L_{\rm X}^{0.6}\propto$ \.m$^{0.6}$
\newline\newline
Essentially, $L_{\rm OIR}$ for BHXBs and NSXBs respond to $L_{\rm X}$ in the same way, but not to \.m, since $L_{\rm X}$ varies with \.m differently for BHXBs and NSXBs. From equations (1) and (7) of \cite{kordet06} we can estimate accretion rates directly from $L_{\rm OIR}$ in the hard state:

\smallskip BHXBs: $L_{\rm OIR}/erg$ $s^{-1}\approx 5.3\times 10^{13}$ (\.m/$g$ $s^{-1}$)$^{1.2}$

\smallskip Or: \.m/$g$ $s^{-1}\approx 3.7\times 10^{-12}$($L_{\rm OIR}/erg$ $s^{-1}$)$^{0.8}$

\smallskip NSXBs: $L_{\rm OIR}/erg$ $s^{-1}\approx 3.2\times 10^{23}$ (\.m/$g$ $s^{-1}$)$^{0.6}$

\smallskip Or: \.m/$g$ $s^{-1}\approx 6.7\times 10^{-40}$($L_{\rm OIR}/erg$ $s^{-1}$)$^{1.7}$
\newline\newline
Given the level of the scatter in the correlations, we expect these calculations to be accurate to $\sim$ one order of magnitude. The OIR luminosity is all that is required to estimate \.m for hard state objects, however \.m would be more accurately measured from $L_{\rm X}$, where most of the energy is usually released. In addition, it is possible to estimate the X-ray, OIR and radio luminosities in the hard state quasi-simultaneously, given the value of just one because they are all linked through correlations.
\newline\newline
\textbf{Parameters of an X-ray Binary:}\newline
Quasi-simultaneous OIR and X-ray luminosities can constrain the nature of the compact object (BHXB or NSXB), the mass of the companion (HMXB or LMXB) and the distance and reddening to an X-ray binary. This is possible because the data from BHXBs (in the hard and soft states), NSXBs and HMXBs lie in different areas of the $L_{\rm X}$--$L_{\rm OIR}$ diagram, with some areas of overlap. If the distance and reddening towards a source is known (not necessarily at a high level of accuracy), its quasi-simultaneous OIR and X-ray fluxes could reveal the source to be any one of the above types of XB. In addition, if the nature of the compact object is known (BH or NS), but the distance and/or reddening is not, the fluxes can constrain these parameters. We stress that the total errors associated with the data (top left error bars in each panel of Figs 1 and 2) need to be considered to define the areas of overlap.

Current techniques used to infer the nature of the compact object in XBs include X-ray timing analysis (e.g. thermonuclear instabilities on accreting neutron stars produce Type I X-Ray bursts), the X-ray spectrum \citep[the well-known X-ray states of BHXBs and tracks in the colour--colour diagrams of NSXBs; e.g.][]{mcclet06} and optical timing analysis in quiescence (the orbital period and radial velocity amplitude constrain the mass function). This new tool has the power to constrain the nature of the compact object requiring only $L_{\rm OIR}$, $L_{\rm X}$ and, only at high luminosities, the X-ray state of the source at the time of observations.

This tool may have many applications. X-ray all-sky monitors such as the \emph{RXTE ASM} are continuously discovering new XBs which are subsequently identified at optical wavelengths. In addition, campaigns are underway to find extragalactic XBs, many of which have optical counterparts discovered with the $HST$ \citep[e.g.][]{willet05}. It would also be interesting to see where ULXs and SMBHs \cite*[e.g. the X-ray and NIR flares seen from Sagittarius A*; e.g.][]{yuanet04} lie in the $L_{\rm OIR}$--$L_{\rm X}$ diagram with the inclusion of a mass term, and whether this could be used to constrain the BH mass in these systems.
\newline\newline
\textbf{The Power of the Jets:}\newline
The evidence in this paper confirms that in the literature of the presence of a flat jet spectrum from radio to NIR wavelengths in the hard state, with the optically thin--optically thick turnover close to the $K$-band (see Sections 3.1 and 3.3). The power of the jets is sensitive to the position of the turnover since it is dominated by the higher energy photons. However, the most recent calculations of the jet power \citep[e.g.][]{gallet05,heingr05,miglfe06} already assume the jet spectrum extends to the IR so no extra calculations are required from this work.

\section{Conclusions}

We have collected a wealth of OIR and X-ray fluxes from 33 XBs in order to identify the mechanisms responsible for the OIR emission. A strong correlation between quasi-simultaneous OIR and X-ray luminosities has been discovered for BHXBs in the hard state: $L_{\rm OIR}\propto L_{\rm X}^{0.61\pm0.02}$. This correlation holds over 8 orders of magnitude in $L_{\rm X}$ and includes data from BHXBs in quiescence and at large distances (LMC and M31), which were \citep[until recently; see][]{gallet06} unattainable for radio--X-ray correlations in XBs. All the NIR (and some of the optical) BHXB luminosities are suppressed in the soft state; a behaviour indicative of synchrotron emission from the jets at high luminosities in the hard state. A similar correlation is found for NSXBs in the hard state: $L_{\rm OIR}\propto L_{\rm X}^{0.63\pm0.04}$, which holds over 7 orders of magnitude in $L_{\rm X}$. At a given X-ray luminosity, a NSXB is typically 20 times fainter in OIR than a BHXB.

Comparing the hard state OIR data to radio data of \cite{gallet03}, we find that the radio--OIR jet spectrum in BHXBs is $\sim$ flat ($F_{\nu}=$ constant) at a given $L_{\rm X}$ \citep[see also][]{fend01}. From this and OIR SEDs of 15 BHXBs, which show an IR excess in a number of systems, we deduce that the optically thick--optically thin turnover in the jet spectrum is likely to be close to the $K$-band for BHXBs. In comparison, the turnover probably lies further into the IR for NSXBs \citep[see][]{miglet06}.

By comparing the observed OIR--X-ray relations with those expected from models of a number of emission processes, we are able to constrain the mean OIR contributions of these processes for XBs. Table 6 summarises the results. We find from the level of soft state quenching in BHXBs that the jets are contributing $\sim$90 percent of the NIR emission at high luminosities in the hard state. The optical BHXB data could have a jet contribution between zero and 76 percent but the optical SEDs show a thermal spectrum indicating X-ray reprocessing in the disc dominates in this regime. In BHXBs, ambiguity arises from the fact that the slope of the expected OIR--X-ray relations from the jets and X-ray reprocessing are essentially indistinguishable. Emission from the viscously heated disc may contribute at low luminosities in BHXBs, but cannot account for the observed correlations. In the NSXBs, the correlations can be explained by the X-ray reprocessing model alone (agreeing with the current thinking), but the jets may play a role at high luminosities, especially in the NIR. The exact contributions of the emission processes are likely to be sensitive to many individual parameters, such as the size of the accretion disc and the shape of the jet spectrum (e.g. the OIR luminosity of the jets is very sensitive to the slope of the optically thick jet spectrum).

\begin{table}
\small
\caption{The OIR emission processes that can describe the empirical OIR--X-ray relations and SEDs.}
\begin{tabular}{p{1.8cm}p{0.8cm}p{0.9cm}p{0.7cm}p{0.8cm}p{0.9cm}}
\hline
Sample&X-ray state&X-ray reprocessing&Jet emission&Viscous disc&Intrinsic companion\\
\hline
NSXBs; OPT&hard&$\surd$&$\times$&$\surd$&$\times$\\
NSXBs; NIR&hard&$\surd$&$\surd$&$\times$&$\times$\\
BHXBs; OPT&hard&$\surd$&$\surd$&$\times$&$\times$\\
BHXBs; NIR&hard&$\times$&$\surd$&$\times$&$\times$\\
BHXBs; OIR&soft&$\surd$&$\times$&$\surd$&$\times$\\
HMXBs; OIR&all &$\times$&$\times$&$\times$&$\surd$\\
\hline
\end{tabular}
\normalsize
\end{table}

The SEDs show a non-thermal component ($\alpha < 0$) in most sources in the hard state, and thermal emission ($\alpha > 0$) is present, probably in all sources in both hard and soft states. The soft state could originate from a combination of X-ray reprocessing (as the OIR--X-ray relations suggest) and the viscous disc \citep[e.g.][]{homaet05}. The SEDs of many BHXBs are redder at low luminosities, which seems to be due to both a cooler disc blackbody and a higher fractional contribution from the jets. BHXBs may be jet-dominated at low luminosities (for more on jet-dominated states, see \citealt*{fend01,falcet04}; \citealt{fendet04,gallet05,kordet06}). We have also made a prediction of the `$L_{\rm OIR}$--$L_{\rm X}$ path' of a BHXB for a typical outburst, based on a hysteresis effect whereby the hard state rise reaches a higher luminosity than the hard state decline, if the source enters the soft state. The data obtained in this paper appear to agree with the prediction, with some inevitable scatter.

Since the X-ray, OIR and radio luminosities and the mass accretion rate are all linked through correlations in the hard state, it is possible to estimate the quasi-simultaneous values of all of these parameters, given the value of just one, by e.g. daily monitoring of the X-ray or OIR fluxes (which is currently being done for some sources by the \emph{RXTE ASM} and by ground-based telescopes). In addition, we have discovered a potentially powerful tool to complement current techniques, that can constrain the nature of the compact object, the mass of the companion and the distance/reddening towards an XB, given only the quasi-simultaneous X-ray and OIR luminosities. Data from BHXBs (in the hard and soft states), NSXBs and HMXBs lie in different areas of the $L_{\rm X}$--$L_{\rm OIR}$ diagram, with small areas of overlap. The tool is most useful for e.g. faint sources with poor timing analysis or at large distances, i.e. extragalactic XBs and new transients.

Further work that could constrain the emission process contributions include identifying linear polarimetry perpendicular to the jet axis (a diagnostic of optically thin synchrotron emission), emission line equivalent width analysis \citep[in particular testing for the Baldwin effect; e.g.][]{mushfe84} and analysing $L_{\rm OIR}$--$L_{\rm X}$ relations in the intermediate and very high X-ray states and in ULXs and SMBHs.

\vspace{5mm}
\emph{Acknowledgements}.
We would like to thank Erik Kuulkers for providing the A0620-00 data \citep{kuul98} and Elena Gallo for the radio and X-ray data of \cite{gallet03}. We thank the referee for many useful comments that significantly improved this work. Results provided by the ASM/RXTE teams at MIT and at the RXTE SOF and GOF at NASA's GSFC. This paper uses data taken with UKIRT and the Liverpool Telescope, the latter of which is funded via EU, PPARC and JMU grants.

\end{document}